\documentclass{amsart}
  \usepackage{amscd,amssymb,epsfig}
   \usepackage{epic,eepic}

 \newcommand{\mapright}[1]{\smash{\mathop{\longrightarrow}\limits^{#1}}}

                     \numberwithin{equation}{subsection}

                     \newtheorem{propo}{Proposition}[subsection]
                     
                     \newtheorem{theor}[propo]{Theorem}
                     \newtheorem{lemma}[propo]{Lemma}
                     \theoremstyle{definition}

                     \theoremstyle{remark}

		     \newcommand{\CC}{\mathbb{C}}

                     \newcommand{\RR}{\mathbb{R}}

		     \newcommand{\Hr}{\mathcal{H}}

		     \newcommand{\str}{\mathcal{S}}

                     \newcommand{\Hom}{\operatorname{Hom}}

\newcommand{\arr}{\operatorname{arr}}

\newcommand{\C}{\mathcal{L}}
\newcommand{\T}{\mathcal{T}}
\newcommand{\W}{\mathcal{W}}
\newcommand{\K}{\mathcal{D}}

		     \newcommand{\pr}{\operatorname{pr}}
		      
		      \newcommand{\card}{\operatorname{card}}

                     \newcommand{\id}{\operatorname{id}}

		     \newcommand{\sign} {\operatorname {sign}}
		    \newcommand{\modu}{\operatorname{mod}}

 \newcommand{\edg}{\operatorname{edg}}
		  \newcommand{\Perm}{\operatorname{Perm}}

\begin{document}
      \title{Loops on surfaces,  Feynman diagrams, and trees}
                     \author[Vladimir Turaev]{Vladimir Turaev}
  
\address{%
              IRMA, Universit\'e Louis  Pasteur - C.N.R.S., \newline
\indent  7 rue Ren\'e Descartes \newline
                     \indent F-67084 Strasbourg \newline
                     \indent France  \newline
		     \indent e-mail: turaev@math.u-strasbg.fr}
                     \begin{abstract} 	 We relate the author's Lie cobracket 
		     in the module additively generated by loops on a surface with the Connes-Kreimer Lie bracket in the module additively generated by trees.
                     \end{abstract}
                     \maketitle

  \section {Introduction}
  
In 1989 the author introduced for any oriented surface $\Sigma$, a Lie cobracket $\nu$ in 
the module $Z=Z(\Sigma)$ generated by the homotopy classes of loops  on $\Sigma$, see \cite{tu1}, \cite{tu}. 
The cobracket $\nu$ complements the Goldman Lie bracket in $Z$ and          
 makes $Z$ into a
Lie bialgebra in the sense of   Drinfeld.
One of the main results of \cite{tu} is   an  algebraic  quantization of this Lie
bialgebra in terms of a Hopf algebra  of knots in $\Sigma\times \RR$. The Goldman Lie bracket has
 a transparent geometric
nature: it is a reformulation of the Poisson bracket determined by the symplectic  structure on the Teichm\"uller
space (and/or other similar moduli spaces). On the other hand, the geometric nature of the cobracket 
$\nu$ remained   mysterious. We argue here that the  
 reason for the existence of this cobracket  
is that   generic loops  on $\Sigma$ can be viewed as Feynman diagrams (of a rather  special type). More
precisely,  we relate   $\nu$  
  to  the  pre-Lie algebras and Hopf algebras of rooted trees introduced by A. Connes  and D. Kreimer \cite{ck} in their fundamental work 
on the algebraic foundations of the 
perturbative quantum field theory. We introduce  similar   algebras generated by  generic loops on   $\Sigma$ and define their canonical projection  
  to the  Connes-Kreimer algebras.

Sections \ref{cob} -- \ref{dcofofcoacd}     deal with  the Lie and pre-Lie (co)algebras arising in this theory.  Section \ref{Hopf} is devoted to  the associated Hopf algebras. In the last Section \ref{furthi1} we consider 
  similar algebraic structures in the realms of Wilson loops and knot diagrams on $\Sigma$.

 Throughout the paper, we fix a
commutative ring
 with unit $R$. The symbol $\otimes$ denotes the tensor product of 
$R$-modules over $R$.  The symbol $\Sigma$ denotes a smooth oriented surface  (possibly with boundary).

  \section{Lie and pre-Lie coalgebras}\label{cob}

 We recall   the algebraic language of pre-Lie and Lie coalgebras used 
 systematically in the paper.

 	          \subsection{Lie   coalgebras}\label{su1}   For an $R$-module $L$, 
denote by $\Perm_L$ the permutation
$x\otimes
y\mapsto y\otimes x$ in $L^{\otimes 2}=L\otimes  L$ and by $\tau_L$ the
permutation $x\otimes y\otimes z\mapsto z\otimes x\otimes y$ in
$L^{\otimes
3}=L\otimes L \otimes   L$.
 A Lie algebra over $R$ can be defined as an $R$-module $L$ endowed 
with an $R$-homomorphism (the
Lie bracket)  $\theta:L^{\otimes 2}\to L$ such that
$\theta \circ
\Perm_L=-\theta$   and  (the Jacobi identity)  $$\theta \circ 
(\id_L\otimes
\theta)
\circ (\id_{L^{\otimes 3}}+\tau_L+\tau_L^2)=0 \in \Hom_R(L^{\otimes 
3},L). $$
   Here  for a set $S$ we denote by $\id_S$   the identity mapping $S\to S$.
 
  Dually, a
{\it Lie coalgebra} over $R$ is an $R$-module
$A$
endowed with an $R$-homomorphism (the Lie cobracket)  $\nu:A\to
A^{\otimes 2}  $
such that   $  \Perm_A \circ \nu=-\nu$   and
			 \begin{equation}\label{jac}(\id_{A^{\otimes
3}}+\tau_A+\tau_A^2)  \circ (\id_A\otimes \nu) \circ \nu=0 \in 
\Hom_R(A,  A^{\otimes 3}).
\end{equation}
A Lie
coalgebra $(A,\nu)$ gives rise to the {\it dual Lie algebra}
  $A^*=\Hom_R(A,R)$ where the Lie bracket   is  the homomorphism  $ 
A^*\otimes A^* \to
A^*$
adjoint  to $\nu$.  For $a,b\in A^*$, the bracket $[a,b]\in A^*$ 
evaluates on     $x\in A$  by
 \begin{equation}\label{jcccc}[a,b] (x)=\sum_i a(x^{(1)}_i) \, b 
(x^{(2)}_i)\in R\end{equation}
for any (finite) expansion $\nu(x)=\sum_i x^{(1)}_i \otimes 
x^{(2)}_i\in A\otimes A$.

A {\it Lie coalgebra homomorphism} $ (A,\nu)\to  (A', \nu')$
is an $R$-linear homomorphism $f:A\to A'$ such that $\nu' f= (f\otimes f)\nu$. 
The adjoint map $f^*:(A')^*\to A^*$ is then a Lie algebra homomorphism.

   \subsection{Pre-Lie algebras and coalgebras}\label{su2} Pre-Lie algebras were 
introduced by Gerstenhaber \cite{ge} and Vinberg \cite{vi} 
independently.
A (right) pre-Lie  algebra  over $R$  is  an $R$-module $L$ endowed 
with an $R$-bilinear multiplication  $L\times L \to L$, denoted
$\star$, such that for any
$x,y,z\in L$,
	 \begin{equation}\label{jacla} (x\star  y) \star  z-  x \star 
(y \star  z)  =  (y\star  x) \star  z- y \star  (x \star 
z)  .\end{equation}
There is a similar notion of left pre-Lie algebras. We will consider 
only right pre-Lie algebras and refer to them simply as pre-Lie 
algebras.
Equality \ref{jacla} implies that
$[x,y]=x\star y -y
\star x$ is a Lie bracket in
$L$. 

A {\it pre-Lie  algebra homomorphism} $(L,\star)\to (L',\star')$
is an $R$-linear homomorphism $f:L\to L'$ such that $ f(x\star y)=  f(x)\star' f(y)$ for all $x,y\in L$.

 Dualizing   Formula  \ref{jacla}, we obtain a notion of    a 
pre-Lie coalgebra. A (left) {\it pre-Lie coalgebra}
is an $R$-module $A$ endowed with an $R$-linear  homomorphism $\rho: 
A\to   A \otimes A$ such that
  \begin{equation}\label{prejacla}(\id_{A^{\otimes 3}}-P^{1,2}_A)\, ((\rho \otimes \id_A)\rho -  (\id_A 
\otimes \rho) \rho)  )=0 \in \Hom_R(A, A^{\otimes 3})\end{equation}
where   $P^{1,2}_A$ is the
 endomorphism of $ {A^{\otimes 3}}$ permuting the first and second  
tensor factors.
Given a pre-Lie coalgebra $(A,\rho)$, the dual module 
$A^*=\Hom_R(A,R)$ acquires a structure of a pre-Lie algebra:
for  $a,b \in A^*$, the value of $a\star  b \in A^*$ on     $x\in A$ 
is given by the right-hand side of Formula \ref{jcccc} for any   expansion $\rho(x)=\sum_i x^{(1)}_i \otimes 
x^{(2)}_i\in A\otimes A$.

\begin{lemma}\label{cor11345} For any pre-Lie coalgebra $(A, \rho)$, the homomorphism  $  \nu=\rho- \Perm_A \rho:A\to A^{\otimes 2}$ is a Lie 
cobracket.
\end{lemma}  
 \begin{proof} It is obvious that   $  \Perm_A \circ \nu=-\nu$.   Formula   \ref{jac}  follows from the identity
 $$(\id_{A^{\otimes
3}}+\tau_A+\tau_A^2)  \circ (\id_A\otimes \nu) \circ \nu=-
 (\id_{A^{\otimes
3}}+\tau_A+\tau_A^2)\circ  (\id_{A^{\otimes 3}}-P^{1,2}_A) \circ  ( (\rho \otimes \id_A) \rho- (\id_A 
\otimes \rho) \rho )$$
which holds for {\it any} $R$-linear homomorphism
$\rho:A\to A^{\otimes 2}$ and $  \nu=\rho- \Perm_A \rho$.
\end{proof}

A {\it pre-Lie coalgebra homomorphism} $(A,\rho)\to  (A', \rho')$
is an $R$-linear homomorphism $f:A\to A'$ such that $\rho' f= (f\otimes f)\rho$. 
The adjoint map $f^*:(A')^*\to A^*$ is then a pre-Lie algebra homomorphism. 
The reader should always keep in mind that a pre-Lie coalgebra homomorphism  $f$ is always a Lie coalgebra homomorphism of the associated  Lie coalgebras and its adjoint $f^*$  is a  Lie  algebra homomorphism
of the dual Lie algebras.

   \section{Pre-Lie   coalgebra  of loops}\label{coac}

 We define  a pre-Lie   coalgebra  of loops on  a smooth oriented surface $\Sigma$.

 	          \subsection{Loops and Feynman diagrams}\label{modic} A {\it generic  loop} on $\Sigma$ is a smooth immersion $\alpha: S^1\to \Sigma-\partial \Sigma$ 
having only
double transversal self-crossings. The set of the self-crossings of $\alpha$ is denoted 
$\# \alpha$; it is always finite. 
We will consider only generic loops  and refer to them simply as 
loops.   The circle $S^1=\{z\in \CC\,\vert \,  \vert z\vert =1\}$ is 
oriented counterclockwise, this makes all  
loops oriented.   
A {\it pointed loop} is a pair (a loop $\alpha$ on $ \Sigma$, a   point    $\ast_\alpha\in \alpha(S^1)- \#\alpha$). The latter point is called the {\it base point} of $\alpha$. 

Two pointed loops $\alpha, \beta$ on $ \Sigma$ can be obtained from each 
other by {\it reparametrization} if $ \ast_\alpha =\ast_\beta $ and there is an orientation preserving
homeomorphism $f:S^1\to S^1$ such that $\beta=\alpha f$.  Pointed loops
$\alpha,
\beta$ on $ \Sigma$ are {\it ambient isotopic} if there is a continuous family of 
homeomorphisms $\{h_t: \Sigma\to \Sigma\}_{t\in [0,1]}$  such that
$h_0=\id_\Sigma$, $\beta=h_1\alpha$, and $\ast_\beta=h_1(\ast_\alpha)$.  We say that two pointed loops  on $ 
\Sigma$ are {\it isotopic} if they can be obtained from each other by 
ambient isotopy and/or reparametrization.  For example, slightly pushing $\ast_\alpha$ along $\alpha(S^1)-\#\alpha$   we obtain a pointed loop isotopic to $(\alpha,\ast_\alpha)$.
It is clear that isotopy of loops is  an equivalence relation. 
We shall
usually identify   loops with their isotopy classes.

 We shall  also  consider loops which are only piecewise smooth. This should create no problem since   the non-smooth points (looking like  corners of a broken line) will be finite in number and distinct from crossing points. All such loops can be   smoothed in the obvious way.

Any loop $\alpha $ on $ \Sigma$ gives rise to a Feynman diagram. It is formed by the circle $S^1$   and a set of straight segments $\{e_p\}_{p\in \#\alpha}$. The endpoints of $e_p$ are the two (distinct) points of the set $\alpha^{-1}(p)\subset S^1$. The segments $\{e_p\}_{p}$  lie in the unit disk 
$D^2=\{z\in \CC\,\vert\, \vert z\vert \leq 1\}$ and have distinct endpoints but  may meet  inside $D^2$. We say that      $p,q\in \#\alpha$ 
		 are {\it linked} if   $e_p\cap e_q\neq \emptyset$ and {\it unlinked} if    $e_p\cap e_q= \emptyset$. We shall see below that the segments $\{e_p\}_{p}$ can be naturally oriented.  The Feynman diagram   ($S^1$, $\{e_p\}_{p}$) is well known in knot theory as the {\it Gauss diagram} of $\alpha$.  In the physical language, the circle $S^1$ represents a fermion and the segments $\{e_p\}_{p}$ represent photons. Note that in this picture the photons do not interact with each other.

   \subsection{Pre-Lie comultiplication for loops}\label{cobra}   Let   $\C=\C(\Sigma)$  be  the   $R$-module freely generated by the set of isotopy classes of   pointed loops  
on $\Sigma$.  Elements of $\C$ are finite formal linear combinations of such isotopy classes with coefficients in $R$.   We define   a  pre-Lie comultiplication
 $\rho:\C \to \C^{\otimes 2}$.

We shall use the following notation: for two
		 distinct points $P,Q\in S^1$, denote by $PQ$ the oriented embedded arc in $S^1$ which starts at $P$, 
		 goes in the positive
(counterclockwise) direction and terminates at  $Q$. Clearly, $PQ\cup QP=S^1$ and $PQ\cap QP=\{P,Q\}$.
For a loop $\alpha:S^1\to \Sigma$,  denote by $\alpha_{P,Q}$ the path in $\Sigma$ 
		 obtained by restricting $\alpha$ to the arc $PQ$.

It suffices to define $\rho:\C \to \C^{\otimes 2}$ on the basis   of $\C$.
Consider the generator of $\C$ presented by a pointed loop  ($\alpha:S^1\to \Sigma$, $\ast_\alpha\in \alpha(S^1)$). This loop traverses every point 
$p\in \# \alpha$ twice in two  different tangent directions.
 The set $\alpha^{-1}(p)\subset S^1$ consists
of  two   points $p_1, p_2\in S^1$  numerated so that starting at $\alpha^{-1}(\ast_\alpha)\in S^1$ and moving along $S^1$ counterclockwise we first meet $p_1$ and then $p_2$.
The path $\alpha_{p_1,p_2}$ (resp.
		  $\alpha_{p_2,p_1}$) 
is  a closed  loop  on $\Sigma$  which starts off 
at $p$ in one of the two tangent directions mentioned above and follows along $\alpha$ until the
first
return to
$p$. The loop $\alpha_{p_2,p_1}$ goes through $\ast_\alpha$ and we take   $\ast_\alpha$ as its base point. As the base point of $\alpha_{p_1,p_2}$ we take $p$.   Set $\varepsilon_p=+1$ if the pair
(the positive tangent direction of $\alpha$ at $p_1$, the positive tangent direction of $\alpha$ at $p_2$) is positive with respect to the orientation of $\Sigma$. In the opposite case set   $\varepsilon_p=-1$. Note that $\varepsilon_p$ and the numeration $p_1,p_2$ depend  on the choice of   $ \ast_\alpha$.  Finally, set
  \begin{equation}\label{minuss}\rho (\alpha,\ast_\alpha)=\sum_{p\in \# \alpha} \varepsilon_p \,   (\alpha_{p_1,p_2},p)    \otimes 
  (\alpha_{p_2,p_1}, \ast_\alpha) .\end{equation}

 \begin{lemma}\label{colbra} The homomorphism $\rho:\C \to 
\C^{\otimes 2}$ is a   pre-Lie comultiplication. 
\end{lemma}
                     \begin{proof} We   verify Formula  \ref{prejacla} for $A=\C$.   Pick a pointed loop 
		     $\alpha$ on $ \Sigma$. 
		A direct application of the definitions shows that both
		    $(\rho \otimes \id)\rho (\alpha)$    and $(\id\otimes \rho)\rho (\alpha)$ 	 are sums of certain expressions numerated by   pairs of unlinked  crossings   $p,q\in \#\alpha$. 
		 Pick such a pair $p,q$. Starting at $\alpha^{-1}(\ast_\alpha)$
		  and moving along $S^1$
		 counterclockwise we   meet the   points $p_1,p_2,q_1,q_2$ in a certain order such that 	 $p_1$ appears before $p_2$ and $q_1$ appears before $q_2$. Exchanging if necessary the letters $p$ and $ q$, 
		 we can assume that the first point we meet is   $p_1$. Then the order in question is either  (i) $p_1,p_2,q_1,q_2$ or (ii) $p_1,q_1,q_2,p_2$. The contribution of $p,q$ to $\rho (\alpha)$ 
		 	is $ \varepsilon_p \, (\alpha_{p_1,p_2},p)    \otimes 
  (\alpha_{p_2,p_1}, \ast_\alpha)+ \varepsilon_q \, (\alpha_{q_1,q_2},q)  
    \otimes  (\alpha_{q_2,q_1}, \ast_\alpha) $.  
		In the case (i) the contribution  of $p,q$ to $(\rho \otimes \id)\rho (\alpha)$  is $0$ and 	   the contribution  of $p,q$ to $(\id\otimes \rho)\rho (\alpha)$ is
		$$\varepsilon_p\varepsilon_q (\alpha_{p_1,p_2},p)
		\otimes  (\alpha_{q_1,q_2},q)\otimes ( \alpha_{p_2,q_1} \alpha_{q_2, p_1}, \ast_\alpha) +
	\varepsilon_p\varepsilon_q (\alpha_{q_1,q_2},q) \otimes  (\alpha_{p_1,p_2},p)\otimes
	(  \alpha_{q_2, p_1} \alpha_{p_2,q_1}, \ast_\alpha).$$ 
	Here  $\alpha_{p_2,q_1} \alpha_{q_2, p_1}$ and $\alpha_{q_2, p_1} \alpha_{p_2,q_1}$  are the loops
	 obtained as   products of the paths  $\alpha_{p_2,q_1}$ and $\alpha_{q_2, p_1} $.
	 Note that up to reparametrization $
	 \alpha_{p_2,q_1} \alpha_{q_2, p_1}=\alpha_{q_2, p_1} \alpha_{p_2,q_1}$. 
	 
	In the case (ii) the contributions of $p,q$ to  $(\rho \otimes \id)\rho (\alpha)$    and $(\id\otimes \rho)\rho (\alpha)$  are both equal to 
	$$\varepsilon_p\varepsilon_q (\alpha_{q_1,q_2},q)   \otimes
	(\alpha_{p_1,q_1} \alpha_{q_2, p_2} , p)\otimes (\alpha_{p_2,p_1}, \ast_\alpha).$$
In both cases  the contribution  of the pair $p,q$ to $((\rho \otimes \id)\rho
-(\id \otimes \rho)   \rho ) (\alpha)$ is invariant under   the permutation of the first and second  tensor factors $P^{1,2}$ and is therefore annihilated by $\id-P^{1,2}$.  This proves  (\ref{prejacla}).
\end{proof}

   The pre-Lie 
comultiplication  $\rho $    induces by
 antisymmetrization a Lie cobracket $\nu$  in $\C$. 
On the basis of $\C$,
\begin{equation}\label{oplk}\nu (\alpha,\ast_\alpha)
=\sum_{p\in \# \alpha} \varepsilon_p \, \left
  (  (\alpha_{p_1,p_2}, p)\otimes (\alpha_{p_2,p_1},\ast_\alpha)-(\alpha_{p_2,p_1},
   \ast_\alpha)\otimes (\alpha_{p_1,p_2},p) \right ).  \end{equation}
 The resulting Lie coalgebra and the dual Lie algebra were introduced in \cite{tu1}, \cite{tu}, cf. Section \ref{dcofofcoacd}.

     \section{Connes-Kreimer  pre-Lie   coalgebras}\label{cokribis}

 We recall (in a convenient form)   and generalize    the definitions of Connes and Kreimer.

 	          \subsection{Pre-Lie   coalgebra of rooted trees}\label{cokr1} 
Connes and Kreimer \cite {ck}  introduced a Lie algebra additively generated by rooted trees.
They observed that the Lie bracket in this Lie algebra is obtained  by antisymmetrization of a pre-Lie 	multiplication. 	  We    describe here the dual pre-Lie comultiplication.

By a {\it tree} we    mean a finite tree.  The set of edges of a tree $T$ is denoted  $\edg(T)$.  A tree $T$ with a distinguished vertex is {\it rooted}, the distinguished vertex being the  
 {\it  root} of $T$.  
A {\it homeomorphism} of rooted trees  is  a homeomorphism  of trees  mapping  
vertices onto vertices,
  edges onto edges, and   the root into the root. Let $\T$ be the $R$-module  
 freely generated by the set of homeomorphism classes of rooted trees.  

Given an edge $e$ of a rooted tree $T$ we obtain two other rooted trees as follows.
Removing (the interior of) $e$ from $T$   we obtain a  
 graph $T-e$ having the same  vertices as $T$.
This graph consists of two disjoint trees $T^1_e$ and $T^2_e$   numerated so that
the root of $T$ lies on $T^2_e$. The root of $T$ provides   a root  for $T^2_e$. As the root  of $T^1_e$ we take  the only vertex of $T^1_e$ adjacent to $e$ in $T$. 

   \begin{lemma}\label{evrcokr11}  The   formula $ \rho (T)=\sum_{e\in \edg (T )}  T^1_e\otimes T^2_e $  defines    a   pre-Lie comultiplication   $\rho:\T\to \T^{\otimes 2}$.
\end{lemma}  

We shall prove a more general statement in the next subsection.

  Antisymmetrizing   $\rho$ we obtain a Lie cobracket $\nu:\T \to \T^{\otimes 2}$. 
 Dualizing  $\rho$ and $\nu$ we obtain   a pre-Lie multiplication $\star$  and a Lie
 bracket in $\T^*=\Hom_R(\T,R)$. Note that the $R$-module $\T$ is based and therefore can be identified with the submodule of $\T^*$
consisting of those $R$-linear functionals $\T\to R$ which are non-zero   only on a finite set of 
(homeomorphism classes of) rooted trees. More precisely, a rooted tree
 $T$ is identified with the functional $ \T\to R$  
taking value 1 on $T$ and value $0$ on all other elements of the basis.
It is easy to check that $\T\star \T\subset \T\subset \T^*$ so that $\T$ acquires the structure of  a pre-Lie 
algebra. This structure and the associated Lie bracket 
  were first defined in \cite{ck}.

   \subsection{Further pre-Lie   coalgebras of trees}\label{cokrext} 
     The     pre-Lie coalgebra $\T$ can be generalized
using   various additional structures on   trees. 
     We describe a general setting for such generalizations.  
     
  By a {\it subtree} of a tree $T$, we mean a tree $T'\subset T$ formed by a set of vertices and edges of $T$.      If $T$ is   rooted then $T'$
 has a unique vertex $v$ such that any path from the root of $T$ to a point of $T'$ passes through $v$. We   take this $v$ as the root of $T'$. In this way all subtrees of a rooted tree become rooted. 

We define a category $RTrees$ whose objects are rooted trees and whose morphisms
 are embeddings. An {\it embedding} of rooted trees $j:T'\to T$ is a homeomorphism of $T'$ onto a subtree of $T$. (Such $j$   sends vertices, edges, and the root of $T'$ onto vertices, edges, and the root of the subtree $j(T')\subset T$.)
For example,    homeomorphisms of rooted trees are embeddings. 
 
  A {\it rooted tree-structure} is a contravariant functor
 from the category $RTrees$ to the category of sets. Such a functor $\varphi$ assigns to any rooted tree $T$ a set $\varphi(T)$ and to any embedding $j:T'\to T$
 a map $\varphi(j):\varphi(T)\to \varphi(T')$. We must  have $\varphi(\id_T)=\id_{\varphi(T)}$ and $\varphi(jj')= \varphi(j') \varphi(j)$ for any   embeddings $j':T''\to T'$ and $ j:T'\to T$.
 We give   two examples of a rooted tree-structure.
 
 (a) (a labeling).  The set $\varphi(T)$   consists of all labelings of vertices and edges of   $T$ by elements of certain sets $S_0$ and 
 $S_1$, respectively. The map 
 $\varphi(j):\varphi(T)\to \varphi(T')$  is   the obvious restriction of labelings.
 
 (b) (a planar structure). The set $\varphi(T)$   consists of all topological embeddings $i:T\to \RR^2$ considered up to composition with an orientation preserving homeomorphism $\RR^2\to \RR^2$. The map 
 $\varphi(j):\varphi(T)\to \varphi(T')$  
 is defined by   $\varphi(j) (i)= ij$.
 
Given a  rooted tree-structure $\varphi$, we define a {\it   rooted $\varphi$-tree}
to be a pair (a rooted tree $T$, an element $s\in \varphi (T)$). Two such pairs $(T,s)$, $(\tilde T, \tilde s)$ are   homeomorphic if 
 there is a homeomorphism $j: T\to \tilde T$ such that $\varphi(j) (\tilde s)=s$.
 For a  subtree $T'\subset T$, set $s\vert_{T'}=\varphi(j) (s)\in \varphi(T')$ where $j$ is the  embedding  $T'\hookrightarrow T$.
 
For any rooted tree-structure $\varphi$, we   define   a {\it (generalized) Connes-Kreimer pre-Lie   coalgebra} $ (\T(\varphi), \rho )$. Here  $ \T(\varphi)$ is the   $R$-module  
 freely generated by the set of homeomorphism classes of  rooted $\varphi$-trees.   (To emphasize the dependence of $R$ we shall sometimes denote this module by  $ \T(\varphi;R)$.)

   \begin{lemma}\label{oplcokr11}  The following formula   defines    a   pre-Lie comultiplication $\rho$ in
     $\T(\varphi)$:
\begin{equation} \rho (T,s)=\sum_{e\in \edg (T )}  (T^1_e, s\vert_{T^1_e})\otimes   (T^2_e,  s\vert_{T^2_e}).  \end{equation} 
\end{lemma}  
 \begin{proof} We must verify Equation \ref{prejacla} for $A=\T(\varphi)$. 
Pick a 	generator $(T,s)$ of $\T(\varphi)$.	A direct application of definitions shows that both    $(\rho \otimes \id)\rho (T,s)$ and $(\id\otimes \rho)\rho (T,s)$  are sums of certain expressions numerated by   pairs    $e_1,e_2\in \edg (T)$. 
		 Pick such a pair $e_1,e_2$. The complement of the (interiors of) these two edges  in $T$ consists of 3 disjoint subtrees $T_0,T_1,T_2\subset T$ where the notation is chosen so that $e_k$ connects a vertex of $T_0$ with a vertex of $T_k$ for $k=1,2$. Set $a_k=(T_k, s\vert_{T_k})\in \T(\varphi)$ where $k=0,1,2$. We   consider three cases depending on whether the root $v$ of $T$ lies in $T_0$, $T_1$, or $T_2$. If $v\in T_0$, then the contributions of the pair $e_1,e_2$ to   
		 $(\rho \otimes \id)\rho (T,s)$ and $(\id\otimes \rho)\rho (T,s)$  are
 respectively $0$ and $a_1\otimes a_2\otimes a_0
 +a_2\otimes a_1\otimes a_0$.
 If $v\in T_1$, then the contributions of the pair $e_1,e_2$ to both  $(\rho \otimes \id)\rho (T,s)$ and $(\id\otimes \rho)\rho (T,s)$   are
 equal to  $ a_2\otimes  a_0\otimes a_{1} $. The case $v\in T_2$ is similar. In all   cases, the contribution  of $e_1,e_2$ to $ ((\rho \otimes \id) \rho -
 (\id \otimes \rho) \rho ) (T,s)$ is invariant under  
		 the permutation of the first  two tensor factors.
	 This gives (\ref{prejacla}).
 \end{proof}
 
  Antisymmetrizing   $\rho$ we obtain a Lie cobracket $\nu$ in $\T(\varphi)$. 
 Dualizing  $\rho$ and $\nu$ we obtain a pre-Lie multiplication $\star$ and a Lie  bracket in $\T(\varphi)^*$. 
  Forgetting the tree-structure yields 
 a pre-Lie (and Lie) coalgebra homomorphism $\T(\varphi)\to \T$ and the adjoint  pre-Lie (and Lie) algebra homomorphism $\T^*\to \T(\varphi)^*$.

We can identify   $\T(\varphi)$   with the submodule of $\T(\varphi)^*$
consisting of   functionals   which are non-zero   only on a finite set of 
(homeomorphism classes of) rooted $\varphi$-trees.  If   the set $\varphi(T)$ is finite for any   $T$, then
  $\T(\varphi)\star \T(\varphi)\subset \T(\varphi)\subset \T(\varphi)^*$ so that $\T(\varphi)$ acquires the structure of  a pre-Lie (and Lie)
algebra. The forgetting  homomorphism $\T^*\to \T(\varphi)^*$ induces then a pre-Lie (and Lie)
algebra homomorphism from  $\T\subset \T^*$ to $\T(\varphi)\subset \T(\varphi)^*$.
It sends a rooted tree $T$ to $\sum_{s\in \varphi(T)} (T,s)$.
 
  Lemma \ref{evrcokr11} follows  from Lemma \ref{oplcokr11} by taking as $\varphi$    the   tree-structure assigning a 1-element set  to every rooted tree.

     \section{From loops to trees}\label{cokrfromus}

 	          \subsection{Homomorphism $\eta$}\label{homt}  We construct  a canonical pre-Lie coalgebra homomorphism  $\eta:\C \to \T(\Phi)$   for an appropriate rooted tree-structure $\Phi=\Phi_\Sigma$. Here  $\Sigma$ is an  oriented surface,  $\C=\C(\Sigma)$,  and $\Phi$ is    a 
		  combination of  a labeling and a planar structure. For a rooted tree $T$, the set $\Phi(T)$ consists of the triples (a labeling of the edges of $T$ by $\pm 1$, a labeling of the vertices of $T$ by  isotopy classes of (non-pointed)  loops on  $ \Sigma $, a planar structure on $T$). In other words, the module $\T(\Phi)$ is generated by  planar rooted trees whose edges are labeled with a sign and    whose vertices are labeled with   loops on $\Sigma$.    Note that forgetting some (or all) of these additional structures on   rooted trees we obtain homomorphisms from $\C$ to other Connes-Kreimer pre-Lie coalgebras. 
		   
		   The definition of $\eta$ goes as follows.    Pick a   loop 
		     $ \alpha:S^1\to \Sigma$. 
		In Section \ref{coac}      we associated with every  crossing  $p\in \#\alpha$    a     segment  $e_p\subset D^2$ with   endpoints
		    on  $ S^1=\partial D^2$. We call a subset $H$ of $\#\alpha$ a {\it cut} of $\alpha$  if $e_p\cap e_q=\emptyset$ for all distinct $p,q\in H$. For a cut $H\subset \#\alpha$, we   write $H\prec\alpha$. With each such $H$  we associate a rooted  $\Phi$-tree $T_H$  as follows. 
		      The segments $\{e_p\}_{p\in H}$  are mutually disjoint  and split the unit disk $D^2$ into several convex  regions called  {\it $H$-faces}.   The vertices of $T_H$ are numerated by the $H$-faces.    The edges of $T_H$ are numerated by elements of $H$: the edge corresponding to $p\in H$ is denoted $[p]$ and connects the      vertices of $T_H$   corresponding to two $H$-faces  adjacent to $e_p$. The graph $T_H$ is dual to the splitting of $D^2$ into the $H$-faces. This graph   can be embedded in  $D^2$ as follows: each vertex is mapped into a  point inside the corresponding $H$-face; each edge is mapped onto the straight segment connecting the images of its endpoints. It is clear from this description that $T_H$ is a planar tree.

	A vertex of $T_H$ arising from an  $H$-face $V$ is labeled with  the  loop on $\Sigma$ obtained as follows:     moving along $\partial V$ we apply $\alpha$ while we are on  	  $\partial V\cap \partial D^2$. The key observation    is that for all $p\in H$, the mapping  $\alpha:S^1\to \Sigma$   maps the endpoints of   $e_p$ to one and the same point, this ensures that our procedure gives a   loop on $\Sigma$ (well defined up to reparametrization). 
		  
It remains to provide $T_H$ with a root and to assign signs to the edges of $T_H$. It is here that we need to assume that $\alpha$  is pointed with base point		  $\ast_\alpha\in \alpha(S^1)$. As the          root of $T_H$ we take the vertex corresponding to the only $H$-face  whose boundary  contains the point $\alpha^{-1}(\ast_\alpha)$.  (We call this $H$-face the {\it root face}.)
 We label each  edge $[p]$ of $T_H$ with  the sign  $\varepsilon_p $ defined in Section \ref{cobra} and set $\varepsilon_H=\prod_{p\in H} \varepsilon_p$.
   The resulting rooted $\Phi$-tree    is denoted   $T_H$ or   $T_H(\alpha,\ast_\alpha)$. Set
\begin{equation}\label{foissy}\eta (\alpha)=\sum_{H\prec\alpha}   \varepsilon_H\,  T_H \in \T(\Phi)\end{equation}
where $H$ runs over all cuts of $ \alpha$. This extends by $R$-linearity to 
a homomorphism  $\eta:\C \to \T(\Phi)$.

   \begin{theor}\label{thettale}   $\eta$ is a pre-Lie coalgebra homomorphism.
\end{theor}  
 \begin{proof}  Pick a pointed loop 
		     $(\alpha:S^1\to \Sigma, \ast_\alpha\in \alpha(S^1))$.   Then
		    $$(\eta \otimes \eta) (\rho (\alpha))=
		    (\eta \otimes \eta) (\sum_{p\in \# \alpha} \varepsilon_p \, (\alpha_{p_1,p_2}, p)\otimes (\alpha_{p_2,p_1},\ast_\alpha) )$$
		    $$=\sum_{p\in \# \alpha}  
\sum_{H_1\prec\alpha_{p_1,p_2} }  \sum_{H_2\prec\alpha_{p_2,p_1} } \varepsilon_p\, \varepsilon_{H_1} \varepsilon_{H_2}  \, 	T_{H_1} (\alpha_{p_1,p_2}, p) \otimes    	T_{H_2}(\alpha_{p_2,p_1},\ast_\alpha). $$ 
 Also 
 $$\rho\eta (\alpha)= \rho(\sum_{H\prec\alpha} \varepsilon_H\,  T_H)$$
 $$ =\sum_{H\prec\alpha}   \varepsilon_H \sum_{e\in \edg ( T_H )}  (T_H)^1_e\otimes (T_H)^2_e  $$
 $$=\sum_{H\prec\alpha}  \varepsilon_H \sum_{p\in H}  (T_H)^1_{[p]}\otimes (T_H)^2_{[p]} $$
   $$=\sum_{p\in \# \alpha} \sum_{p\in H\prec\alpha}   \varepsilon_H  (T_H)^1_{[p]}\otimes (T_H)^2_{[p]}.$$  Therefore it suffices to prove that for every $p\in \# \alpha$,   
$$\sum_{H_1\prec\alpha_{p_1,p_2} }  \sum_{H_2\prec\alpha_{p_2,p_1} } 
  \varepsilon_p\, \varepsilon_{H_1} \varepsilon_{H_2}  \,	T_{H_1} (\alpha_{p_1,p_2}, p) \otimes    	T_{H_2}(\alpha_{p_2,p_1},\ast_\alpha)$$
$$=\sum_{p\in H\prec\alpha}      \varepsilon_H (T_H)^1_{[p]}\otimes (T_H)^2_{[p]}.$$ 
This equality  follows from the existence of the bijective  correspondence $(H_1,H_2) \mapsto \{p\}\cup H_1\cup H_2$
  between   pairs  of cuts
 ${H_1 \prec \alpha_{p_1,p_2} } , {H_2 \prec \alpha_{p_2,p_1} }$
 and  cuts $H\prec \alpha$ containing $p$. 
Under this correspondence, $\varepsilon_p \,\varepsilon_{H_1} \varepsilon_{H_2}=\varepsilon_H$, 
 $T_{H_1} (\alpha_{p_1,p_2}, p)=(T_H)^1_{[p]}$, and $T_{H_2}(\alpha_{p_2,p_1},\ast_\alpha)=
 (T_H)^2_{[p]}$.
 \end{proof} 
 
  \subsection{Remarks}\label{rem1} 1. Each loop on $\Sigma$ has a natural degree   defined as the number of its self-intersections. This   can be used to define  an   $R$-linear homomorphism $r:\T(\Phi;R)\to \T(\Phi; R[u])$ where $R[u]$ is the ring of 1-variable polynomials with coefficients in $R$. The homomorphism $r$ sends
  a   rooted $\Phi$-tree $T$ to $u^{\vert T\vert} T$ where 
  $\vert T\vert$ is the total degree of $T$ defined as the sum over the vertices of $T$ of the degrees of the corresponding loops. It is clear that $r$ is a pre-Lie coalgebra homomorphism
  and so is $r \eta:\C\to \T(\Phi; R[u])$. The latter homomorphism allows us to separate  the terms of different total degrees in the expression for $\eta(\alpha)$. Quotienting $r \eta$ by $u$
  we obtain a pre-Lie coalgebra homomorphism $r \eta (\modu u):\C\to \T(\Phi; R )$
  which is given by the same formula  as $\eta$ but with $H$ running over all  cuts of $\alpha$ such that  $\vert T_H\vert =0$.   The equality $\vert T_H\vert =0$ means that   the loops labeling the vertices of $T_H$ have no self-crossings. This can be rephrased by saying that    $H$ is a {\it maximal} cut of $\#\alpha$  not contained in a bigger cut.

2.   In the definition of $\eta$,  the tree-structure $\Phi$ can be lifted to a stronger tree-structure $\tilde \Phi$.  Observe that the edges of a planar tree incident to a vertex $v$ are cyclically ordered. A pair of consecutive edges is called a {\it corner} at $v$. The tree-structure $\tilde \Phi$ is formed by   $\Phi$ and a choice of the corner 
at the root of the tree. For a subtree $ T'\subset  T$, 
the restriction mapping   $ \tilde \Phi(T)\to  \tilde \Phi(T')$
is defined as follows. If $T'$ contains  the   root $v$ of $T$ then  the distinguished corner of $T'$ at $v$ is the one that contains the distinguished corner of $T$ at $v$. If $T'$ does not contain   the   root   of $T$ then  we distinguish  the corner at the root of $T'$ containing the only edge $e$ of $T$ such that $T'=T^2_e$. The trees $T_H$ above all have a distinguished corner at the root, namely the corner containing the point $\alpha^{-1}(\ast_\alpha)$. This   lifts   $\eta$ to a pre-Lie coalgebra homomorphism 
$\C \to \T(\tilde \Phi)$.

3. The comultiplication $\rho$ can be included in a   family  of pre-Lie comultiplications $\rho^{a,b}$ in $\C$ parametrized by pairs $a,b \in R$. To define $\rho^{a,b}$, we simply replace 
 $\varepsilon_p$ in the definition of $\rho$ by $a+b \varepsilon_p$. Replacing $\varepsilon_H$ by 
 $\prod_{p\in H} (a+b \varepsilon_p)$ in the definition of $\eta$ we obtain a pre-Lie coalgebra homomorphism from $(\C,\rho^{a,b})$ to $\T(\Phi)$.  For $b=0$, the  
 pre-Lie coalgebra $(\C,\rho^{a,b})$  is independent of the choice of orientation in $\Sigma$ and can be defined for non-orientable surfaces. 
 
 4.  Cuts on loops were   introduced in \cite{tu}, Section 15  where they are used   to relate loops on $\Sigma$ to  knots in $\Sigma\times \RR$. 
 
 5. Non-generic loops on $\Sigma$ also lead to  interesting  and quite involved algebraic structures. The author plans to study them elsewhere.

 6. The work of  Chas and Sullivan \cite{cs} suggests that the constructions of this paper generalize to loops in manifolds of arbitrary dimension.
   
   \section{Weaker Lie   coalgebras}\label{cofofcoac}

In this section we  address the following question: can one define algebraic coproducts as above under weaker assumptions on loops and trees? Specifically, we are interested in  non-pointed loops and  oriented but  non-rooted trees. The pre-Lie comultiplications defined above do not survive in this setting. However, as we show here,  the associated Lie cobrackets do survive. 

 	          \subsection{Loops re-examined}\label{modad}
	Forgetting the base points in the definition of isotopy of loops on an   oriented surface $\Sigma$, we obtain   isotopy for (non-pointed) loops. 	   Denote by $\C_0$   the   $R$-module freely generated by the set of isotopy classes of (non-pointed) loops 
on $\Sigma$.  Forgetting the base point yields a   projection $\pr:\C\to \C_0$.

\begin{lemma}\label{colbra35} The Lie cobracket 
 $\nu$ in $\C$ induces a Lie cobracket  $\nu_0$ in $\C_0$. 
\end{lemma}
                     \begin{proof} We need   to prove that when we forget the base points of loops 
		     on the right-hand side of Formula \ref{oplk}, 
		    the resulting expression does not depend on the choice of       $\ast_\alpha$.   The reason for this comes from the fact that for each $p\in \# \alpha$,  the two   points of the set $\alpha^{-1}(p)$ have a natural order  $p^1,p^2$  independent of   $\ast_\alpha$.  This order  is defined by the condition that the pair (the positive tangent direction of $\alpha$ at $p^1$, the positive tangent direction of $\alpha$ at $p^2$) is positive with respect to the orientation of $\Sigma$. If $\alpha$ is pointed then 
		       $p^1=p_1, p^2=p_2$ in the case $\varepsilon_p=1$ and $p^1=p_2, p^2=p_1$ in the case $\varepsilon_p=-1$. In all cases
$$\varepsilon_p \, (    \alpha_{p_1,p_2} \otimes  \alpha_{p_2,p_1}-\alpha_{p_2,p_1} \otimes  \alpha_{p_1,p_2}   )=    \alpha_{p^1,p^2} \otimes  \alpha_{p^2,p^1}-
\alpha_{p^2,p^1} \otimes  \alpha_{p^1,p^2}   .$$
We can  thus write down an explicit formula for $\nu_0$:
\begin{equation}\label{klikl} \nu_0(\alpha)=\sum_{p\in \#\alpha}\alpha_{p^1,p^2} \otimes  \alpha_{p^2,p^1}-
\alpha_{p^2,p^1} \otimes  \alpha_{p^1,p^2}.\end{equation}
\end{proof}

 	          \subsection{Lie   coalgebra of oriented trees}\label{cokr2} 
An {\it oriented tree} is a   tree with   oriented edges. Any subtree of an oriented tree is   oriented in the obvious way. We define a category $OTrees$ whose objects are oriented trees and whose morphisms
 are orientation preserving embeddings (mapping vertices to vertices and edges to edges).   An {\it oriented tree-structure} is a contravariant functor $\psi$
 from the category $OTrees$ to the category of sets.
Given an oriented tree-structure $\psi$,   an  {\it  oriented $\psi$-tree}
is a pair (an oriented tree $T$, an element $t\in \psi (T)$). Two such pairs $(T,t)$, $(\tilde T, \tilde t)$ are   {\it homeomorphic} if 
 there is a homeomorphism $j: T\to \tilde T$ such that $\psi(j) (\tilde t)=t$.

For any oriented tree-structure $\psi$, we   define an $R$-module  $ \T_0(\psi)$
 freely generated by the set of homeomorphism classes of  oriented $\psi$-trees. 
Removing an edge $e$ from an oriented tree $T$ we obtain two  disjoint  subtrees   $T^1_e, T^2_e\subset T$   numerated so that
$e$ is directed from a vertex of $T^2_e$ to a vertex of $T^1_e$.  

   \begin{lemma}\label{cobbbb11}  For any oriented tree-structure $\psi$, the following formula  defines    a    Lie cobracket in $ \T_0(\psi)$:
$$\nu_0 (T,t)=\sum_{e\in \edg (T )} (T^1_e, t\vert_{T^1_e})\otimes   (T^2_e,  t\vert_{T^2_e}) - (T^2_e, t\vert_{T^2_e})\otimes   (T^1_e, t\vert_{T^1_e}) . $$
\end{lemma} 

This is proven along the same lines as Lemma \ref{oplcokr11}; the difference is  that instead of various
 positions of the root   one has  to consider four possible orientations on $e_1,e_2$. (The identity used in the proof
 of Lemma \ref{cor11345} and a similar identity with $P^{1,2}$  replaced by $P^{2,3}$ may  help  to shorten the computations.) Warning: the homomorphism $ \T_0(\psi)\to \T_0(\psi)^{\otimes 2}$ defined by $(T,t)\mapsto \sum_{e } (T^1_e, t\vert_{T^1_e})\otimes   (T^2_e,  t\vert_{T^2_e})$ is {\it not} a pre-Lie cobracket.  
 
 If the set $\psi(T)$ is finite for all $T$, then the Lie cobracket $\nu_0$ induces a Lie bracket in  $ \T_0(\psi)$ using the  standard embedding $\T_0(\psi)\hookrightarrow \T_0(\psi)^*$.

  Every rooted tree admits a canonical orientation  uniquely defined by the condition that all edges adjacent to the root  are outgoing and   all other vertices are
 adjacent to exactly one incoming edge. This defines  a covariant functor $h:RTrees \to OTrees$.

We shall be particularly interested in the oriented   tree-structure $\Psi=\Psi_\Sigma$ assigning to an oriented tree $T$  the set    of   pairs (a labeling of the  vertices of $T$ by isotopy classes of (non-pointed)  loops on  $ \Sigma $, a planar structure on   $ T $).  Let $\Phi=\Phi_\Sigma$ be the rooted tree-structure defined in Section \ref{homt}.     For a  rooted tree $T$ and   $s\in \Phi(T)$, let $\sign_s(T)$ be the product of the signs labeling the   edges of $T$.
Let $h_s(T)$ be the oriented   tree    obtained from $h(T)$ by inverting orientation on all edges    labeled with $ -1$.   The $\Phi$-structure $s$ induces a $\Psi$-structure $s'$ on $h_s(T)$   by keeping the labels of the vertices and the embedding into $\RR^2$. 
 We define   an $R$-linear homomorphism
$\pr_{\T}:\T(\Phi)\to \T_0(\Psi)$ by $\pr_{\T}(T,s)=  \sign_s(T)  (h_s(T),s')$.   

  \begin{lemma}\label{hobbit11}  The homomorphism $\pr_{\T}:\T(\Phi)\to \T_0(\Psi)$  is a   Lie coalgebra homomorphism.
\end{lemma}

The proof is an exercise on the definitions.

   \subsection{Homomorphism $\eta_0$}\label{homtrerere}  We define a  Lie coalgebra homomorphism  $\eta_0:\C_0 \to \T_0(\Psi)$ (a version of $\eta$ for non-pointed loops).     For a   loop 
		     $ \alpha:S^1\to \Sigma$, set
   $$\eta_0 (\alpha)=\sum_{H\prec\alpha}     T_H \in \T_0(\Psi) $$
where $T_H$ is the planar (non-rooted) tree determined by   $H$. The labels of the vertices of $T_H$ are   as in Section \ref{homt}. The edges of $T_H$ are oriented as follows. For a crossing $p\in \#\alpha$, we orient the segment $e_p\subset D^2$ from $p^1$ to $p^2$ (in the notation introduced in the proof of Lemma \ref{colbra35}) and orient the edge $[p]\subset T_H\subset D^2$ so that the pair   ($[p]$, $e_p$) determines the counterclockwise orientation of $D^2$. (By the definition of $[p]$,   it intersects $e_p$ transversally in one point.) The next lemma follows directly from the definitions.

    \begin{lemma}\label{newthetr}       The following diagram  is commutative: 
   $$
\begin{array}{ccc}
\C&\mapright{\pr}&\C_0\cr
\eta\downarrow&&\downarrow\eta_0\cr
\T(\Phi)&\mapright{\pr_{\T}}&\T_0(\Psi).
\end{array}
$$
\end{lemma}

    \begin{theor}\label{newcorororo}      $\eta_0$ is a Lie coalgebra homomorphism.
\end{theor}  
 
 \begin{proof} By the results above   $\eta$ and $\pr_{\T} $ are Lie coalgebra homomorphisms. Hence so is $\eta_0\circ \pr= \pr_{\T}\circ \eta: \C \to \T_0(\Psi)$. Since $\pr:\C \to \C_0$ is a surjection, $\eta_0$ is a Lie coalgebra homomorphism. \end{proof}  
 
 Remark \ref{rem1}.1 applies in this setting
 with obvious changes.
  
  \subsection{Related pre-Lie and Lie coalgebras}\label{relcobra} The constructions above can be adapted to
so-called virtual strings, see \cite{tu3}.
An {\it open} (resp. {\it closed)  virtual string  of rank
$n$} is a subset of $]0,1[$ (resp. of $S^1$) consisting of $2n$ distinct points partitioned into $n$  ordered  pairs.
These pairs  are called   {\it arrows}.
The set of arrows of a virtual string $a$ is denoted   $\arr(a)$.
Two open (resp. closed) virtual strings $a, b$ are {\it homeomorphic} if there is an   orientation preserving  self-homeomorphism of $[0,1]$ (resp. of $S^1$)
transforming $a$ into $b$.

Pick an arrow $e$ of an open virtual string $a$ with endpoints $p_1,p_2\in ]0,1[$ numerated so that $p_1<p_2$. Set $\varepsilon_e=1$ if $e$ is directed from $p_1$ to $p_2$ and $\varepsilon_e=-1$ otherwise. Denote  by 
$a^1_e$ (resp. $a^2_e$) the virtual string obtained from $a$ by removing $e$ and all other arrows with at least one endpoint on $]0,1[- [p_1,p_2]$
 (resp. on $[p_1,p_2]$). The formula 
$$ \rho (a)= \sum_{e\in \arr(a)}\varepsilon_e a^1_e\otimes a^2_e$$
defines a pre-Lie comultiplication in the   $R$-module $\str$ freely generated by the set of homeomorphism classes of open virtual
strings. This comultiplication is  connected with   obvious multiplication in $\str$ given by   concatenation of open strings. Namely,
$\rho (ab)=\rho(a) (1\otimes b)+ (1\otimes a)\rho (b)$ for   $a,b\in \str$.

Closed virtual strings can be obtained from the open ones by gluing $0$ and $1$. This gives a   projection from  $\str$ to the similar module $\str_0$  generated by homeomorphism classes of closed virtual strings. The pre-Lie comultiplication does not survive this operation but the associated Lie cobracket   survives. The homomorphisms $\eta$ and $\eta_0$ have their analogues: a 
pre-Lie coalgebra homomorphism   $\str\to \T(\Phi')$ and a 
 Lie coalgebra homomorphism   $\str_0\to \T_0(\Psi')$ where $\Phi'$ is a rooted tree-structure combining a labeling  of edges by $\pm 1$ with a planar  structure and  $\Psi'$ is a   planar  structure.

Finally, observe that there is a projection from   the coalgebras of loops on   an oriented  surface $\Sigma$ into the   coalgebras of virtual strings.  The key observation is that every pointed loop $\alpha$ on    $\Sigma$  determines an open virtual string $a(\alpha)$ formed by the ordered pairs $(p^1,p^2)$
with  $p\in \#\alpha$. Here we identify $S^1-\alpha^{-1}(\ast_\alpha)$ with $]0,1[$ via an orientation preserving homeomorphism.   The formula $\alpha\mapsto a(\alpha)$ defines a pre-Lie coalgebra homomorphism
$\C (\Sigma)\to \str$.  It is in general neither surjective nor injective. In particular, pointed loops on $\Sigma$
related by the action of the mapping class group   have the same images in $\str$. The homomorphism
$\C (\Sigma)\to \str$ induces a Lie coalgebra homomorphism
$\C_0 (\Sigma)\to \str_0$. We also have the obvious forgetting homomorphisms $\T(\Phi)\to \T(\Phi')$ and $\T_0(\Psi)\to \T_0(\Psi')$ making all the natural diagrams arising here commutative.

  \section{Lie   bialgebra of loops}\label{dcofofcoacd}    
     
    We relate the Lie coalgebra $\C_0=\C_0(\Sigma)$ to the Lie bialgebra  of loops on   $\Sigma$ introduced in \cite{tu1}, \cite{tu}.  
    
      \subsection{Lie coalgebra $Z=Z(\Sigma)$}\label{modulez}   Loops $\alpha, \beta$ on $ \Sigma$ are {\it freely homotopic} if there is a mapping $f:S^1\times [0,1]\to \Sigma$ such that $\alpha(a)=f(a,0) $ and $ \beta (a)=f(a,1)$ for all $a\in S^1$. 
      Free homotopy   is an equivalence relation on the set of loops. The corresponding set of  equivalence classes is denoted   $\hat \pi=\hat \pi (\Sigma)$.   This set   has a distinguished element   $  \alpha_0  $ represented by an embedding $ S^1\hookrightarrow \Sigma$ onto the boundary of a small disk in $\Sigma$. 
       For connected $\Sigma$,  the set  $\hat \pi$  can be identified with the set of conjugacy classes in the fundamental group $\pi$ of $\Sigma$.

  Let $Z $ be the $R$-module freely generated by the set $\hat \pi$.    
Since isotopic loops are   homotopic, assigning to an isotopy class of loops the underlying homotopy class we obtain an $R$-linear homomorphism $P:\C_0\to Z$.  
   The Lie cobracket  $\nu_0$ in $\C_0$ cannot directly induce a Lie cobracket in $Z$ because of the following obstruction. Consider a   loop $\alpha:S^1\to \Sigma$ and insert a small  $\varphi$-like cirl on the right of $\alpha$. This   gives a new loop, $\alpha'$,  homotopic to $\alpha$. It is clear from Formula \ref{klikl}, that $$(P\otimes P) \nu_0(\alpha')=(P\otimes P)\nu (\alpha)+\alpha_0\otimes P(\alpha) -P(\alpha) \otimes \alpha_0\neq (P\otimes P)\nu_0(\alpha).$$
  This obstruction can be circumvent as follows. Let $g:Z\to Z$ be  the  $R$-linear endomorphism  defined by $g(a)=a$ for all $a\in \hat \pi-\{ \alpha_0 \}$ and $g(  \alpha_0  )=0$.
  
 \begin{lemma}\label{vvvvvvlbra} (\cite{tu1}, \cite{tu})  The following formula 
 defines a Lie cobracket $\nu_Z:Z\to Z^{\otimes 2}$: \begin{equation}  \nu_Z(\alpha)=(g\otimes g)\left (\sum_{p\in \#\alpha}\alpha_{p^1,p^2} \otimes  \alpha_{p^2,p^1}-
\alpha_{p^2,p^1} \otimes  \alpha_{p^1,p^2}\right ). \end{equation}
  \end{lemma} 
  
  Formula \ref{klikl} implies that   $g P:\C_0\to Z$ is a Lie coalgebra homomorphism.   The map  $\C_0\to \T_0(\Psi)$ does  not survive the factorization of loops by homotopy: the  linear combination of trees associated with a loop may change  drastically under homotopy. However, there appears another fundamental   structure described next. 
      
  \subsection{Goldman's Lie  bracket in  $Z$}\label{bibi}    
     Goldman \cite{go} defined a  Lie 
bracket  $[,]$ in $Z$   as follows. (A related Lie algebra is implicit in the earlier paper of Wolpert \cite{wo}.)   Let
$\alpha,
\beta $ be two loops on
$\Sigma$.  Applying a small isotopy to $\alpha$ we can assume that 
$\alpha$ meets $\beta$  transversally at a finite number of points
distinct from  self-intersections of $\alpha, \beta$.  Denote
the (finite) set
$\alpha(S^1)
\cap
\beta(S^1)$ by $\alpha \# \beta$.  Each point  $p\in \alpha\#\beta$ 
is a   double transversal
intersection of $\alpha$ and $\beta$.  Let $(\alpha \cdot 
\beta)_p=\pm 1$ denote the intersection index of
$\alpha$ and
$\beta$ at $p$. Smoothing 
the set $\alpha(S^1) \cup
\beta(S^1)$ at $p$ we obtain   a loop on $\Sigma$ denoted $(\alpha 
\beta)_p$. This smoothing replaces the $X$-like crossing at $p$ by two
disjoint arcs $\uparrow \uparrow$ so that arriving  to a neighborhood 
of $p$  along $\alpha$ (resp. $\beta$) one leaves along $\beta$ (resp.
$\alpha$).   Set
	 $$[\alpha, \beta]= \sum_{p\in 
\alpha \# \beta} (\alpha \cdot \beta)_p (\alpha\beta)_p 
.$$
Extending by bilinearity we obtain a bracket $[,]$ in $Z$.

 \begin{theor}\label{lbra} (\cite{go})  $[,]$ is a well defined Lie bracket in 
$Z$. \end{theor}
                    
  To explain the connection between the Lie cobracket $\nu_Z$  and Goldman's Lie bracket, we recall the notion  of a Lie bialgebra due to V. Drinfeld. A {\it Lie bialgebra} over $R$ is an $R$-module  $A$
endowed with a Lie bracket $[,]$ and a Lie cobracket   $\nu:A\to
A^{\otimes 2}  $
such that  
 $\nu ([x,y])=  x \nu (y)-y \nu (x)  
 $ for any $x,y\in A$.
Here $A$ acts on $A\otimes A$ by $x (y\otimes z)= [x,y] \otimes z + y 
\otimes [x,z]$.

 \begin{theor}\label{colbira} (\cite{tu1}, \cite{tu}) The triple $(Z, [,], \nu_Z)$ is a 
 Lie bialgebra. \end{theor}

  This bialgebra has a topological quantization (in fact, a biquantization) in terms of a   Hopf algebra of skein classes of oriented links in $\Sigma\times \RR$. It is curious to note that this algebra acts on the spaces of conformal blocks associated with $\Sigma$ by appropriate 2-dimensional modular functors.
  
   \section{Hopf algebras of trees and loops}\label{Hopf}

 	          \subsection{Symmetric algebras}\label{pol} Given an $R$-module $A$, one   has its symmetric algebra
		  $$S(A)=\oplus_{n\geq 0} S^n (A)$$
		  where $S^0(A)=R$, $S^1(A)=A$, and $S^n (A)$ is the $n$-th symmetric tensor power of $A$ for $n\geq 2$.    The algebra $S(A)$ is commutative and associative and has a unit $1\in R=S^0(A)$. The projection $S(A)\to S^0(A)=R$ along $\oplus_{n\geq 1} S^n (A)$ is called the {\it augmentation}. 
		  If $A$ is a free module with basis $\{x_i\}_{i}$, then $S(A)$ can be identified with the polynomial algebra $R[\{x_i\}_{i}]$.

 \subsection{Connes-Kreimer Hopf algebras}\label{ckha} Consider  the symmetric algebra $S(\T)$  whose elements are polynomials on rooted trees with coefficients in $R$. (The unit $1\in S^0(\T)$ can be thought of as an empty   tree.)  Connes and Kreimer \cite{ck} defined a non-cocommutative comultiplication   in $S(\T)$ which makes  it  into 
 a   bialgebra. We recall   their  definition extending it (in a straightforward way) to the setting of rooted trees with   structure. Fix  a rooted  tree-structure $\varphi$.  A {\it simple cut} of  a rooted tree  $T$ is a    set   $c\subset \edg (T)$ such that any embedded path leading from the root of $T$ 
 to a vertex of $T$   meets at most one element of $c$.   Removing from $T$ all (open) edges belonging 
 to a simple cut $c$ we obtain a   set of   disjoint subtrees of $T$. One of them denoted $T_0 $ contains the root of $T$. The other subtrees $\{T_e\}_{e\in c}$ are numerated by the elements of $c$ so that each $e\in c$ connects a vertex of $T_0$ to a vertex of $T_e$.   Recall that all subtrees of $T$ are rooted in a canonical way.  For $s\in \varphi (T)$, set 
 $$l_c (T,s)=\prod_{e\in c}  (T_e, s\vert_{T_e}) \in   S (\T(\varphi)),\,\,\,\, r_c(T,s)=(T_0, s\vert_{T_0})\in   S^0(\T(\varphi)).  $$
Set
\begin{equation}\label{cokr}\nabla (T,s)=(T,s) \otimes 1   +\sum_{c} l_c(T,s) \otimes  r_c(T,s)   \end{equation}
 where $c$ runs over all simple cuts of $T$.
 Note that the term 
  $l_c  \otimes  r_c $  corresponding to $c=\emptyset$ is equal to $1\otimes (T,s)$. Formula \ref{cokr} defines $\nabla$ on the generators of the algebra $S=S(\T(\varphi))$; it extends uniquely to an algebra homomorphism
 $\nabla:S \to S \otimes S $. 
 It follows from the definitions that the augmentation $ S \to R$ is a counit of  $\nabla$.
 Connes and Kremer proved that $\nabla$ is coassociative. 
 They also explain that the resulting  bialgebra $S(\T(\varphi))$ has an antipode and is thus a Hopf algebra.  
 
 \subsection{Hopf algebra of pointed loops}\label{poilopha} 
 Consider the symmetric algebra $S=S(\C)$  where  $\C=\C (\Sigma)$ is the $R$-module freely generated by isotopy classes of pointed loops on an oriented surface $\Sigma$.  Elements of $S$ are polynomials on  isotopy classes of pointed loops on $\Sigma$ with coefficients in $R$. We define a comultiplication $\Delta$ in $S$ as follows.   Pick a pointed loop $\alpha$ on $ \Sigma $. 
Recall the segments $\{e_p\}_{p\in \#\alpha}$ in the unit disk $D$ and the $H$-faces of $D$ determined by a  cut $H\prec\alpha$, cf. Sections \ref{modic} and \ref{homt}. Let $v(H)$ denote the root $H$-face, i.e.,   the only $H$-face containing $\alpha^{-1} (\ast_\alpha)\in \partial D$.
For    $p\in H$ denote by $v (H,p)$ the   unique $H$-face adjacent to $e_p$ and such that $v  (H,p)$ and     $v(H)$ lie on different sides of  the line containing  $e_p$.  The formula $p\mapsto v (H,p)$ establishes a bijection between   $H$ and the set of $H$-faces distinct from $v(H)$. For any $H$-face $v$ denote by $\alpha_v$ the associated pointed loop on $\Sigma$. Set
$$l_H(\alpha)=\prod_{p\in H} \alpha_{v(H,p)}\in   S,\,\,\,\,r_H(\alpha)=\alpha_{v(H)} \in    \C\subset S.$$

A cut $H$ of $\alpha$ is {\it simple} if  all segments $\{e_p\}_{p\in H}$ are adjacent to $v(H)$. To indicate that $H$ is a simple cut of $ \alpha$ we write $H\ll \alpha$.   Set 
$$\Delta (\alpha)=\alpha \otimes 1   +\sum_{H\ll \alpha}  \varepsilon_H \, l_H(\alpha) \otimes  r_H(\alpha) \in S  \otimes S   $$ 
where $\varepsilon_H=\prod_{p\in H} \varepsilon_p$.
Note that the term  $\varepsilon_H \,l_H(\alpha) \otimes  r_H(\alpha) $  corresponding to $H=\emptyset$ is equal to $1\otimes \alpha$.

This defines $\Delta$ on the generators of $S$; it extends uniquely to an algebra homomorphism
 $\Delta:S \to S \otimes S $.

 \begin{lemma}\label{dfcolbra35}  
 $\Delta$   is coassociative. 
\end{lemma}
                     \begin{proof}  It suffices to prove that $(\id\otimes \Delta)\Delta (\alpha)=(\Delta\otimes \id)\Delta (\alpha)$ for any pointed loop $\alpha$ on $\Sigma$. 
		     Set $$A=(\id\otimes \Delta)\Delta (\alpha)- \Delta(\alpha)\otimes 1,\,\,\,\, B=
	(\Delta\otimes \id)\Delta (\alpha)  - \Delta(\alpha)\otimes 1.$$
	We shall prove that $A=B$. To this end we   define another expression $C$ and prove that $A=C=B$. 
	
For simple cuts  $G\ll \alpha, G'\ll \alpha$ we   write $G'\leq G$ if    $v(G')\subset v(G) $. Then $G\cup G'\subset \#\alpha$ is a   cut of $\alpha$. Its  faces are the $G$-faces $\{v(G\cup G',p)=v(G,p)\}_{p\in G}$ and the faces obtained by splitting $v(G)$ along the segments $\{e_q\}_{q\in G'-G}$, specifically, $\{v(G\cup G',q)\}_{q\in G'-G}$ and   $v(G')$. (Note that $G'-G=G'-(G\cap G')$.) Set 
$$C=\sum_{G, G'\ll \alpha \,s.t.\, G'\leq G} \varepsilon_{G\cup G'}  
\prod_{p\in G} \alpha_{v(G,p)}\otimes \prod_{q\in G'-G} \alpha_{v(G\cup G',q)}\otimes \alpha_{v(G')} \in S^{\otimes 3}.$$

We   claim  that $A=C$.  If follows from the definition of $\Delta$ that
$$
A=\sum_{H\ll \alpha} \varepsilon_H \sum_{H'\ll r_H(\alpha)} \varepsilon_{H'} l_H(\alpha)\otimes l_{H'} (r_H(\alpha))\otimes r_{H'} (r_H(\alpha)).$$
The formula $(G,G')\mapsto ( H=G , H'=G'-G)$ defines 
  a bijective correspondence between   pairs $(G\ll \alpha,  G'\ll \alpha)$ such that $G'\leq G$  and    pairs $(H\ll \alpha, H'\ll r_H(\alpha))$.  The corresponding  terms of $A$ and $C$ are equal.  The equality of signs follows from the formula   $\varepsilon_{G\cup G'}=\varepsilon_{G}\,\varepsilon_{G'-G}=\varepsilon_{H}\varepsilon_{ H'}$. 
  Therefore $A=C$.

We   claim  that $B=C$. If follows from the definition of $\Delta$ that
$$
B=\sum_{H\ll \alpha} \varepsilon_H \prod_{p\in H} \left (\alpha_{v(H,p)}\otimes 1  + \sum_{H_p\ll \alpha_{v(H,p)}} \varepsilon_{H_p} l_{H_p} (\alpha_{v(H,p)}) \otimes r_{H_p} (\alpha_{v(H,p)})\right )
\otimes r_H(\alpha)$$  
$$=\sum_{H\ll \alpha} \varepsilon_H \sum_{I\subset H} \,\, \sum_{\{H_p\ll \alpha_{v(H,p)}\}_{p\in H-I}} \varepsilon_{\cup_p H_p} \left( \prod_{q\in I }  \alpha_{v(H,q)}  \prod_{  p\in H-I}
   l_{H_p} (\alpha_{v(H,p)})\right ) \otimes
\prod_{  p\in H-I} r_{H_p} (\alpha_{v(H,p)}) 
\otimes r_H(\alpha).$$ With each tuple  $(H\ll \alpha, I\subset H, \{H_p\ll \alpha_{v(H,p)}\}_{q\in H-I})$ we associate the pair $
 ( G=I\cup \cup_{p\in H-I} H_p , G'=H)$. This  defines 
  a bijective correspondence between   such tuples  
and   the pairs $(G\ll \alpha, G'\ll \alpha)$ such that $ G'\leq G$. The corresponding  terms  of $B$ and $C$ are equal.  The equality of signs follows from the formula   $ {G\cup G'}=H\cup \cup_{p\in H-I} H_p$ and the fact that the sets   $\{H_p\}_{p\in H-I}$ and $H$ are pairwise disjoint.    Therefore $B=C$. \end{proof}

It is clear that the augmentation $ \varepsilon: S \to R$ is a counit of  $\Delta$. The bialgebra $(S,\Delta)$ has an antipode $s$. This is an algebra endomorphism of $S$  determined on a generator  $\alpha$   by induction on   $\vert \alpha\vert =\card (\#\alpha)$: if $ \vert \alpha\vert=0$, then $s(\alpha)=-\alpha$, if $ \vert \alpha\vert\geq 1$, then
$$s(\alpha)=-\alpha-\sum_{H\ll \alpha, H\neq \emptyset}  \varepsilon_H \,  l_H(\alpha)   \, s(r_H(\alpha)) \in S$$
where we use that $\vert r_H(\alpha)\vert <\vert \alpha\vert$. These formulas guarantee that $m (\id_S\otimes s)\Delta (\alpha) =\varepsilon(\alpha)$ where $m$ is multiplication in $S$. In other words,  
$s$ is a left inverse of $\id_S$ with respect to the convolution product $\star$ in $\Hom_R(S,S)$ defined by $f\star g= m(f\otimes g)\Delta$ for $f,g\in \Hom_R(S,S)$.   Similar inductive formulas show that $\id_S$ has a right inverse
$s'\in \Hom_R(S,S) $ and then $s=s \star (\id_S\star s') =(s \star  \id_S)\star s'=s'$. Therefore $s$ is   an antipode for  $S$.

  \subsection{Homomorphism $ \eta$}\label{homu} The $R$-linear homomorphism  $\eta: \C \to \T(\Phi)$ defined in Section \ref{homt} extends by multiplicativity to an algebra homomorphism 
 $S(\C) \to S(\T(\Phi))$ also denoted $\eta$.

   \begin{theor}\label{newthettale}   $ \eta$ is a Hopf algebra homomorphism.
\end{theor}  
 \begin{proof} We need to show  that $\nabla ( \eta(\alpha)) = ( \eta\otimes  \eta) (\Delta(\alpha))$ for  any pointed loop 
		     $\alpha:S^1\to \Sigma$. 
Set $a=\nabla ( \eta(\alpha))- \eta(\alpha)\otimes 1$ and $b= ( \eta\otimes  \eta) (\Delta(\alpha) - \alpha\otimes 1)$. It is enough  to check that $a=b$.

 Consider a  cut  $ H\subset \# \alpha$ of $\alpha$ and a subset $G\subset H$ such that   $G$ is a simple cut of $\alpha$. 
The cut $G$ determines   a simple cut $c(G,H)$ of the tree $T_H=T_H(\alpha)$ consisting of  the edges 
$\{\langle p\rangle\}_{p\in G}$. (This establishes a bijection between simple cuts of $T_H$ and  subsets of $H$ which are simple cuts of $\alpha$.) Set
$$\langle G,H\rangle=l_{c(G,H)} (T_H)\otimes  r_{c(G,H)}(T_H) \in S(\T(\Phi))\otimes  \T(\Phi).$$
 By abuse of notation we do not specify the $\Phi$-structure  	on 
the trees   on the right-hand side; it is induced by the one on $T_H$. 
It is easy to deduce from the definitions that 
$$ b=\sum_{G\ll \alpha} \varepsilon_G \sum_{ G\subset H\prec \alpha }\varepsilon_{H-G}\, \langle G,H\rangle=\sum_{H\prec \alpha} \varepsilon_H \sum_{G\subset H\, s.t.\, G\ll \alpha}\langle G,H\rangle=a.$$  

 Thus $ \eta$ is a bialgebra homomorphism. Finally, any bialgebra homomorphism of Hopf algebras is a Hopf algebra homomorphism, see \cite{sw}, Lemma 4.0.4. \end{proof}
	     
	 \subsection{Non-commutative Hopf algebra  of loops}\label{ncHaol}
 Given an $R$-module $A$, one   has its tensor algebra
		 $T(A)=\oplus_{n\geq 0} A^{\otimes n}$ 
		  where $A^{\otimes 0} =R$, $A^{\otimes 1} =A$, and $A^{\otimes n}$ with $n\geq 2$ is the tensor product over $R$ of $n$ copies of $A$.  The product in $T(A)$ is defined by $$(a_1\otimes...\otimes a_n)(a_{n+1}\otimes...\otimes a_{n+m})=a_1\otimes... \otimes a_{n+m}$$ for $a_1,..., a_{n+m}\in A$.  
		  In the sequel instead of $a_1\otimes...\otimes a_n$ we  write    $\prod_{i} a_i$.  The algebra $T(A)$ is   associative and has a unit $1\in R=A^{\otimes 0}$. The identity map $A\to A$ extends to a surjective  algebra homomorphism  $T(A)\to S(A)$. 
		  If $A$ is a free module with basis $\{x_i\}_{i}$, then $T(A)$ is the  algebra of non-commutative polynomials in the variables  $\{x_i\}_{i}$ with coefficients in $R$.

 Consider the tensor algebra $T=T(\C)$  where  $\C=\C (\Sigma)$. Observe  that any simple cut $H\subset \#\alpha$ of a pointed loop $\alpha$ is totally ordered in a canonical way. Namely starting at the base point $\ast_\alpha$ and moving along the loop we meet first a certain point of $H$ twice, then another point  of $H$ twice, etc. The resulting  order on $H$ allows us to set 
 $\tilde l_H(\alpha)=\prod_{p\in H} \alpha_{v(H,p)}\in  \C^{\otimes n} \subset T$.
 The   formula $$\tilde \Delta (\alpha)=\alpha \otimes 1   +\sum_{H\ll \alpha}  \varepsilon_H \, \tilde l_H(\alpha) \otimes  r_H(\alpha)    $$ 
defines a map $\C\to T\otimes T$. It extends to an algebra homomorphism
 $\tilde \Delta:T\to T\otimes T$.  The same argument as in the proof of Lemma \ref{dfcolbra35} shows that  
 $\tilde \Delta$   is coassociative.  In this argument in  the expressions for $C,B$ one should use the orders in $G'-G$ and $H-I$ (needed in the second tensor factor)   induced by the  orders in $G'$ and $H$ respectively.   In  the expression for $B$ one should replace $\prod_{q\in I }  \alpha_{v(H,q)}  \prod_{  p\in H-I}
   l_{H_p} (\alpha_{v(H,p)})$ with $\prod_{p\in H} a_p$ where 
   $a_p=\alpha_{v(H,p)}$ for $p\in I$ and $a_p=l_{H_p} (\alpha_{v(H,p)})$ for $p\in H-I$.
   
The projection $T\to R$ along $\oplus_{n\geq 1} \C^{\otimes n}$ is a counit of $T$.
 The existence of an antipode in $T$ is straightforward.   
 It is clear that the natural projection $T\to S(\C)$ is a Hopf algebra homomorphism.

A non-commutative analogue of $\eta$ is a Hopf algebra homomorphism $\tilde \eta$ from $T$ to   Foissy's  \cite{fo} non-commutative Hopf algebra $\Hr_{P,R}(\Phi) $ generated by   rooted $\Phi$-trees. Note that Foissy considers planar rooted trees with labeled vertices but nothing prevents from extending his definitions  to the case where the edges are also labeled. (Alternatively, one may observe that the edges of a rooted tree are numerated by the vertices distinct from the root so that a labeling of the edges can be interpreted as  a labeling of the vertices.) 
The value of   $\tilde \eta$ on the generators of $T=T(\C)$ is given  by Formula \ref{foissy}. We have a commutative diagram of Hopf algebra homomorphisms 
 $$
\begin{array}{ccc}
 T(\C)&\mapright{\tilde \eta}& \Hr_{P,R}(\Phi) \cr
 \downarrow&&\downarrow \cr
S(\C)&\mapright{\eta}&S(\T(\Phi) )
\end{array}
$$
where the vertical arrows are the natural projections.

	 \subsection{Remark}\label{rem2} As in  Remark \ref{rem1}.3, for any  $a,b\in R$, we can   replace everywhere (and in particular in the definition of $\varepsilon_H$) the sign $\varepsilon_p$ with  
 $a+ b \varepsilon_p$. This yields  a 2-parameter family of coassociative comultiplications $\Delta^{a,b}$  in $S(\C)$ (resp. $\tilde \Delta^{a,b}$ in $T(\C)$) and Hopf algebra homomorphisms from the resulting Hopf algebras to $S(\T(\Phi))$
 (resp. to $\Hr_{P,R}(\Phi)$).

		   \section{Further algebras}\label{furthi1}

 In analogy with   tree-structures, we can  introduce   axiomatically certain \lq \lq structures" on loops suitable for a generalization  of the   comultiplications $\rho, \Delta$ defined above. Instead of doing this here, we  focus on  two specific additional structures on loops and briefly discuss the associated algebras.

  \subsection{Algebras of Wilson loops}\label{wils} By a {\it region} of a (generic) loop  $ \alpha:S^1\to  \Sigma$ on an oriented  surface $\Sigma$, we mean a connected component of $\Sigma- \alpha(S^1)$. A {\it Wilson loop}   is a    loop on $  \Sigma$ whose regions are endowed with   numbers (say, real or complex). The number associated with a region is called its {\it area}.   A   Wilson loop is {\it pointed} if its underlying  geometric loop   is pointed. Isotopy of (pointed) Wilson loops is defined in the obvious way, the areas  being preserved under ambient isotopies and  reparametrizations.
  
 For a Wilson loop $\alpha$ and a crossing $p\in \#\alpha$,  both loops $\alpha_{p_1,p_2}$ and $\alpha_{p_2,p_1}$ appearing in   Section \ref{cobra} become  Wilson loops as follows. The area of a region   $X$ of 
$\alpha_{p_1,p_2}$ (resp. of $\alpha_{p_2,p_1}$) is   set   to be the sum of the areas of  regions of $\alpha$ contained in $X$.

 Let $\W$ be the $R$-module freely generated by the set of isotopy classes of pointed Wilson loops. The definition of the pre-Lie comultiplication   in Section \ref{cobra} applies to Wilson loops word for word 
  and gives a pre-Lie comultiplication in $\W$. In analogy with Section \ref{cofofcoac}, the associated Lie cobracket in $\W$ induces a Lie cobracket in $\W_0$, the $R$-module freely generated by the set of isotopy classes of  non-pointed  Wilson loops.      Forgetting the areas   we obtain  a pre-Lie coalgebra homomorphism $\W\to \C$ and  a  Lie coalgebra homomorphism $\W_0\to \C_0$.  Similarly, the 	definition of   $\Delta$    in Section \ref{Hopf}   applies to Wilson loops  
  and gives   Hopf algebra structures in $S(\W)$ and $T(\W)$
  and a commutative diagram 
 $$
\begin{array}{ccc}
 T(\W)&\mapright  {}&T(\C)  \cr
 \downarrow&&\downarrow \cr
S(\W)&\mapright  {}&S(\C )
\end{array}
$$
of surjective Hopf algebra homomorphisms.  The  comultiplications  in $\W, S(\W), T(\W)$ can be included in  a 2-parameter family of comultiplications $\rho^{a,b}, \Delta^{a,b}, \tilde \Delta^{a,b}$ as in Remarks \ref{rem1}.3 and \ref{rem2}.

 \subsection{Algebras of  knot diagrams}\label{knotdm}   A {\it knot diagram}  on  an oriented  surface  $\Sigma$ is a  (generic)  loop   $ \alpha:S^1\to  \Sigma$ such that each crossing $p\in \#\alpha$ is endowed with a sign $\mu_p=\pm 1$. The equivalence with the more standard language of over/undercrossings is established as follows. Recall that  two branches of $\alpha$ passing through $p\in \#\alpha$ have an order determined by the orientation of $\Sigma$ (cf. the proof of Lemma \ref{colbra35}). Then the first branch goes over (resp. under) the second branch if $\mu_p=1$ (resp. if $\mu_p=-1$). Note that by the definition of a loop, our knot diagrams are oriented. 
 
 A   knot diagram is {\it pointed} if its underlying  geometric loop   is pointed. Isotopy of (pointed) knot diagrams is defined in the obvious way, the signs $\mu$  being preserved under ambient isotopies and  reparametrizations.

 Let $\K$ be the $R$-module freely generated by the set of isotopy classes of pointed knot diagrams. 
 Pick four elements $a,b,c,d\in R$.
 For a pointed knot diagram $\alpha$ and a crossing $p\in \#\alpha$,  both loops $\alpha_{p_1,p_2}$ and $\alpha_{p_2,p_1}$ appearing in   Section \ref{cobra} become  knot diagrams: their self-crossings 
 are also self-crossings of $\alpha$ and we attribute to them the same signs $\mu$. Set 
 $$\rho^{a,b,c,d} (\alpha, \ast_\alpha )=\sum_{p\in \# \alpha} (a+ b \varepsilon_p +c\mu_p +d \varepsilon_p  \mu_p)\,    (\alpha_{p_1,p_2} ,p)    \otimes 
  (\alpha_{p_2,p_1}, \ast_\alpha) .$$
 This defines a pre-Lie comultiplication $\rho^{a,b,c,d}:\K\to \K\otimes \K$. 
 Note that multiplying all signs $\mu_p$ by $-1$ we obtain an isomorphism   $(\K, \rho^{a,b,c,d})\approx (\K, \rho^{a,b,-c,-d})$. For $c=d=0$, this defines an involution in  $(\K, \rho^{a,b,0,0})$.

    Similarly, the 	definition of   $\Delta$    in Section \ref{Hopf}   can be applied   to pointed knot diagrams and gives   Hopf algebra comultiplications $\Delta^{a,b,c,d}$  in $S(\K)$ and $\tilde \Delta^{a,b,c,d}$  in $T(\K)$. (It is understood that we   replace everywhere    $\varepsilon_p$ with  
 $a+ b \varepsilon_p +c\mu_p +d \varepsilon_p  \mu_p$.) The definitions of $\eta, \tilde \eta$ also apply   and yield     a pre-Lie coalgebra homomorphism $(\K, \rho^{a,b,c,d})\to \T(\Phi)$ and 
 a commutative diagram of   Hopf algebra homomorphisms
 $$
\begin{array}{ccc}
 (T(\K), \tilde \Delta^{a,b,c,d})&\mapright  {}&\Hr_{P,R}(\Phi)  \cr
 \downarrow&&\downarrow \cr
(S(\K),  \Delta^{a,b,c,d}) &\mapright {}& S(\T(\Phi)).
\end{array}
$$

   For $a=c=0$, the  Lie cobracket in $\K$ associated with $\rho^{a,b,c,d}$ induces a Lie cobracket in $\K_0$, the $R$-module freely generated by the set of isotopy classes of  non-pointed  knot diagrams.
If additionally  $  d=0, b=1$, then  we have a forgetting Lie coalgebra homomorphism $\K_0\to \C_0$.
 
\subsection{Homomorphisms}\label{homcvcvcm} The pre-Lie algebras $\C, \W, \K$ are related by  three pre-Lie algebra homomorphisms:
\begin{equation}\label{drf} (\C, \rho^{a,b}) \longrightarrow (\K, \rho^{a,b,0,0}) \longrightarrow (\W, \rho^{a,b}) \longrightarrow (\C, \rho^{a,b}).\end{equation} 
We describe them on the generators. 
The leftmost homomorphism is obtained by attributing   $\mu =+1$ to all crossings of a pointed loop. The rightmost homomorphism is obtained by forgetting the areas. The  middle homomorphism $ \K  \to  \W $
comes from the  theory of shadow knots  \cite{tun}. It   transforms a pointed knot diagram $(\alpha, \mu)$ into a pointed Wilson loop   as follows. A crossing $p\in \# \alpha$ is adjacent to four (possibly coinciding)  regions $R_1,...,R_4$ of $\alpha$ which we numerate so that $R_1$    lies  between the   outgoing branches of $\alpha$ at $p$ and $R_3$ lies  between the   incoming  branches of $\alpha$ at $p$ while $R_2$, $R_4$ are the two remaining regions. 
Then $p$ contributes $(-1)^{k+1} \mu_p/2$ to the area  of $R_k$ for $k=1,...,4$.
The area of a region of $\alpha$ is defined to be the sum of the   contributions of the crossings of $\alpha$
adjacent to this region.   
The makes $\alpha$ into a  pointed Wilson loop.  

It is clear that the composition of the three homomorphisms in (\ref{drf}) is the identity map. These homomorphisms induce Hopf algebra homomorphisms
$$(S(\C),  \Delta^{a,b})  \longrightarrow (S(\K),  \Delta^{a,b,0,0})  \longrightarrow (S(\W),  \Delta^{a,b}) \longrightarrow (S(\C),  \Delta^{a,b})$$
and  similar Hopf algebra homomorphisms with $S, \Delta$ replaced by $T, \tilde\Delta$.


\begin{thebibliography}{CJKLS}
		     
		     


 \bibitem[CS]{cs} M. Chas, D. Sullivan, \emph{String Topology\/},  math.GT/9911159.
 

 \bibitem[CK]{ck} A. Connes, D. Kreimer, \emph{Hopf algebras, renormalization and noncommutative 
 geometry\/},
 Comm. Math. Phys.  199  (1998),  no. 1, 203--242. 


 \bibitem[Fo]{fo} L. Foissy,  \emph{Les alg\`ebres de Hopf des arbres enracin\'es d\'ecor\'es. I\/}.    Bull. Sci. Math.  126  (2002),  no. 3, 193--239.

\bibitem[Ge]{ge} M. Gerstenhaber,
 \emph{The cohomology structure of an associative ring\/},
Ann. of Math. (2) 78 (1963),  267--288.

  \bibitem[Go]{go} W.   Goldman,
 \emph{The symplectic nature of fundamental groups of surfaces\/},
Adv. Math. 54  (1984), 200--225.

   
 \bibitem[Sw]{sw} M.E. Sweedler, Hopf algebras, W.A. Benjamin, Inc. New York, 1969.
  


 \bibitem[Tu1]{tu1} V. Turaev,
\emph{ Algebras of loops on surfaces, algebras of knots, and quantization\/}, 
 Braid group, knot theory and statistical mechanics, 
 59--95, Adv. Ser. Math. Phys., 9, World Sci. Publishing, Teaneck, NJ, 1989.

 \bibitem[Tu2]{tu} V. Turaev,
\emph{Skein quantization of Poisson algebras of loops
on surfaces\/},
Ann. Sci. \'Ecole Norm. Sup. (4) 24 (1991), no. 6, 635--704.
 
   \bibitem[Tu3]{tun} V. Turaev,
\emph{Shadow links and face models of statistical mechanics\/}, 
J. Differential Geom. 36 (1992), no. 1, 35--74.

   \bibitem[Tu4]{tu3} V. Turaev,
\emph{Virtual strings and their cobordisms\/}, math.GT/0311185. 


\bibitem[Vi]{vi}  E.B. Vinberg,
\emph{The theory of convex homogeneous cones\/},
Trans. Mosc. Math. Soc. 12 (1963), 340--403; translation from Tr. 
Mosk. Mat. Ob-va 12 (1963), 303--358.

\bibitem[Wo]{wo}  S. Wolpert,
\emph{ On the symplectic geometry of deformations of a hyperbolic surface\/},  Ann. of Math. (2)  117  (1983),  no. 2, 207--234.
                     \end{thebibliography}
                     \end{document}